\documentclass[useAMS,usenatbib]{mn2e}
\usepackage{graphicx}
\usepackage{longtable}

\def\be{\begin{equation}}
\def\ee{\end{equation}}

\title[Clustering and the Search for Dim and Dark Galaxies]{Clustering and the Search for Dim and Dark Galaxies}
\author[M.J. Disney. R.H.Lang and J Ott.]
{M.J. Disney$^{1}$\thanks{E-mail: mjdisney@gmail.com (MJD); otheremail@otheraddress (HL)} R.H. Lang$^{1}$\footnotemark[1] and J.Ott$^{2}$\footnotemark[1]
\\
$^{1}$School of Physics and Astronomy, Cardiff University, The Parade, Cardiff, CF24 3AA, Wales, UK.\\ $^{2}$ Pete V. Domenici Science Operations Centre VLA, PO Box 0
1003 Lopezville Rd,Socorro,NM 87801-0387 USA.}

\date{Accepted 1Received 2021; in original form }

\pagerange{\pageref{firstpage}--\pageref{lastpage}} \pubyear{2002}

\begin{document}
\maketitle

\label{firstpage}

\begin{abstract}
We re-investigate the question of whether there is a significant population of Dim (Low Surface Brightness) and/or Dark Galaxies (DDGs). We argue that if they are clustered with bright ones then a physical resolution of $\simeq$ 10 kpc. will be needed to  distinguish their 21-cm. and QSO Absorption Line(QSOAL) counterparts from their brighter neighbours, leading to a real possibility of confusion. But until now such a resolution has not been available in this context. New  Very Large Array (VLA)  observations  reveal that the identifications with bright optical objects claimed in previous single-dish blind HI surveys are often unreliable. For instance  14/36 of our high-resolution (5 arc sec, 13kpc.) sample have no optical counterparts in the Digital Sky Survey. This suggests that DDGs might be commonplace after all, and we go on to re-examine the main arguments that have been used against them.. We find that the QSOAL argument founders on the  aforementioned clustering while deep CCD surveys  have so far covered  too small an area to set set strong constraints on Low Surface Brightness Galaxies (LSBGs). A cosmos filled with low surface brightness galaxies, dark galaxies and intergalactic gas clouds can no longer be ruled out.\end{abstract}


\section{Introduction}
The vast majority of known galaxies have effective surface brightnesses (SBs) which lie in a narrow peak within 2 or 3 magnitudes of the terrestrial sky (roughly 22 magnitudes $arc sec ^{-2}$ i.e. 22$B\mu\ $ or $21V\mu$). And yet, due to special circumstances, some few galaxies are known which span a vast range in intrinsic SB between 13$B\mu\ $ and 30.5 $B\mu\ $ i.e. a spread of ten million. We see the high SB tail of compact galaxies in the Hubble Deep Field where Tolman surface-brightness dimming by $\sim$9 magnitudes, brings compact redshift 7 galaxies down into the �normal� range near 22 $B\mu\ $ (Disney and Lang 2012), and we see the extreme low SB tail in the Local Group where their individual stars can be detected above Galactic star counts (e.g Belukurov et al 2007,  Koposov et al. 2015, Bechtol et al 2015, Beasley et al 2016) down to 30.5$B\mu\ $.  The narrow peak in SB contrasted with the extremely broad range naturally suggests that dramatic Selection Effects are at work. And that wouldn�t be surprising; even outside the atmosphere the combination of scattered (zodiacal) sunlight and scattered starlight means that the background glare in our locality is 10 magnitudes brighter than one would expect it to be in intergalactic space (32 $B\mu\ $) if only the summed intensity of $\emph{known} $ populations of galaxies were responsible. Theoretical analyses of the  surface-brightness selection effects indeed show that the only galaxies with any hope of detection will lie in a very narrow wigwam-shaped peak less than 3 mags wide FWHM centred a couple of mags brighter than the sky ( Disney and Lang 2012).       	Apart from the above special cases there can be no hope of by-passing this  glare in the optical itself (see Discussion and Appendix C below) so more ingenious indirect methods are necessary to unearth the potentially hidden population of dim and dark galaxies (DDGs). For instance using redshift to discriminate between background and local signals � as in the 21 centimetre Hydrogen line, or by using absorption where a galaxy much dimmer than the glare would nevertheless betray its presence through the absorption features it would impress on the spectrum of a background source such as a QSO.	Over the past 20 years strenuous efforts have been made to find DDGs using both the above indirect techniques with, it has to be said, singular lack of success [For instance the HIPASS survey  carried out from Parkes (Zwaan et al 2003) in which we participated, found more than 4000 blind HI sources in the Southern hemisphere $\--$ all of which turned out to have easily visible optical counterparts (Doyle et al 2005).] The trouble with all that negative evidence is that it's  too good to be true, far too good. From time to time, usually by chance, odd observations turn up to refute it. For instance the colossal LSBG Malin 1, 200 kpc. in diameter and containing more than $10^{11}$ solar masses of HI, appeared by chance in the background  (at 25,000 kms) during a modest Arecibo  survey targeted at individual dwarfs in the Virgo Cluster at $\sim$ 1100 km $s^{-1}$ (Bothun et al 1989). Surely it can't be unique $\--$ and yet its analogues have never been found in the aforesaid blind HI surveys of almost the entire sky.  And latterly deep optical observations of the Coma cluster (van Dokkum et al 2014,  Koda T et al 2015)  reveal that it contains many galaxies of Milky Way size but with surface brightnesses up to 5 mags dimmer.   Indeed the failure  of the 21 cm. and QSOALs searches has become so egregious that one begins to suspect that some other explanation  than the lack of the missing galaxies must be responsible. In this paper we claim it is the clustering of the DDGs with the �normal� ones � which leads to frequent  confusions of identity. In the case of HI sources the poor resolution of the single-dish radio telescopes used to carry out the blind HI surveys means that it was all too easy to find a bright optical galaxy at approximately the right redshift to explain the HI signal when in fact a much dimmer galaxy clustered nearby was the true emitter. We back that suspicion up with high resolution observations recently made with the newly enhanced VLA which confirm that the previous optical identifications were often incorrect and that DDGs are instead sometimes the seat of the signals. And in the case of QSOALs, clustering means that DDGs within arc secs of the QSO sight-lines are 30 times more likely to be responsible for the absorptions than the  giant luminous galaxies arc minutes away with hypothetical giant halos hundreds of  kilo-parsecs in extent which are invoked  now (Indeed such halos would appear to be redundant).	

The rest of the paper is arranged by section as follows:
(2) Demonstrates the interplay between clustering and identification and the challenge it poses at both 21 cm  and for QSOALs.
\\
(3) Describes our recent observations with the uprated VLA of HIPASS IDs which show that our suspicions about their reliability are confirmed. Moreover 12/19 of the new blind sources we find could be DDGs.
\\
(4) Extends the discussion to single-dish HI surveys in general to reach the disconcerting conclusion  that increasing dish-size is of no help in solving the identification problem because, in general,  larger dishes find their blind sources a correspondingly further distance away where their $\emph{physical}$  resolution, which is what matters for identification purposes, is no better. We then calculate the identification-Ambiguity in  existing single-dish blind surveys to find that they are all unreliable.
\\
(5) Examines whether the hosts of damped Lyman-alpha QSOALS's can be identified reliably using a combination of HST and ground based 8-Metres. Because of clustering, the majority, including dim and dark hosts, probably cannot. Thus the inferred existence of enormous gaseous halos around luminous spiral galaxies has to be questioned. The QSOALs are more naturally explained as due to  smaller  dim and dark galaxies much closer to the lines of sight.
\\
(6) The Discussion returns to the general conjecture that there is a rich population of Hidden Galaxies still to be found. We examine, the HI, the QSOALS and the deep CCD evidence and conclude that  the many refutations of this conjecture so far published all founder on the difficulty of reliable identifications in a strongly clustered universe. A rich Low Surface Brightness Universe is a real possibility, and  well worth looking for.

 \section{THE PROBABILITY OF SPURIOUS IDENTIFICATIONS}
 It has often been assumed that the probability of spurious optical identifications in the case of 21-cm sources (and QSOALSs) is rather small because one has radial velocity as well as position as a discriminator. This would be true if galaxies were not so heavily clustered. Unfortunately, in the clustered case, the vast majority of plausible misidentifications will be with  other brighter galaxies in the same Group where the velocities are  very similar too. Velocity is therefore not an independent test of association - except on  large scales or where there is high velocity-precision. This will be particularly true of gas-rich galaxies, the majority of which seem to lie in loose groups with low velocity dispersions (eg Pisano et al 2011) so that for velocity to be a useful discriminant the difference between the  HI source velocity  and the optical galaxy velocities (including the errors)  would need to be $\pm$ 15 km sec$^{-1}$or less ( see later) which, in practice,because of the errors, will rarely be the case. So if dim or dark galaxies are clustered with bright ones there is, as we shall show next, every chance that they will be misidentified with a visible one nearby.

 To see this we first carry out a crude clustering calculation. Let $\eta(r)$ be the number of clustered galaxies within a sphere of radius r of some HI source then:
 \be\eta(r)=\overline{n}\int_{0}^{r}4\pi x^{2}\cdot [1+\xi(x)]\cdot dx\ee
 where      \be \xi(x)\equiv(\frac{r_0}{x})^{1.8}\ \ \ \ \           (r_0\simeq 8 Mpc)\ee
 is the 3-dimensional clustering function (e.g. Peebles 1993) which should be accurate for the range of distances relevant here ($\leq 2Mpc.)$. $\overline{n}$ is the number-density of $L_{*}$ galaxies considered, averaged over all space, not just the cluster. Then
 \be\eta(r)=\frac{4\pi\bar{n}}{3}\cdot[r^{3}+106r^{1.2}]\ee

where r is in Mpc and $\bar{n}$ in galaxies $Mpc.^{-3}$ . For distances out from the source of $\leq$  1Mpc.,as are relevant here, the second term is utterly dominant. From Eqn.(3) we generate Table 1 which shows the number of visible galaxies one can expect to encounter at various $\emph{projected}$ distances $\hat{r}$ out from one of them. The second column shows the expected number of $L_{*}$ galaxies, the third column shows the number of all galaxies up to 4 mag. fainter than $L_{*}$ that might be accepted as identifications. The fourth column gives a more realistic estimate of column 3 (see later). We have used the SDSS g-band Luminosity Function (Blanton et al. 2003) for which $\bar{n}\sim 10^{-2}.L_{*} gals.Mpc.^{-3}$. It is clear from the calculation that leads to the last column (below) that $\emph{if}$ dark and dim galaxies are clustered with SDSS ones then at projected distances of only 24 kpc. away from them there is a 50 per cent chance of encountering a visible but spurious identification. This is the severe challenge posed by clustering to the correct optical identification of both HI sources and QSOALS.

TABLE 1 HERE

\begin{table*}
\centering
\begin{minipage}{140mm}
\caption{Number of possible spurious galaxy identifications}
\begin{tabular}{cccc}

r'(kpc)  & n($L_{*}$) & n($L_{*}$+ up to 4mag) &Last col adj for end cylinders
\\
\hline
40 & 0.12 & 0.72 & 1.3
\\
80 & 0.27 & 1.6 & 2.9
\\
160 &0.61 & 3.7 & 7
\\
320 & 1.0 & 6 & 11
\\
\hline
\end{tabular}
\end{minipage}
\end{table*}

 If we invert the calculation and ask how precise a radio position has to be in order to reduce the chance of a false identification $\--$ in the above sense $\--$ to say only 25 per cent per source, the answer is 13 kpc $\emph{at any distance}$. To see how challenging this is, Table 2 enumerates the positional accuracy  demanded of a source in arc min by the  13 kpc. criterion at various radial velocities ($ H_0= 75 \ km\ s^{-1} Mpc^{-1} $ throughout.

TABLE 2 HERE

\begin{table*}
\centering
\begin{minipage}{140mm}
\caption{Precision required to avoid misidents at different distances; M=0.25;q=6.}
\begin{tabular}{cc}
Velocity & $\delta\theta$ (arc min)
\\
\hline
1000 & 3.2
\\
2000 & 1.6
\\
4000 & 0.8
\\
8000 &0.4
\\
16,000 & 0.2
\\
\hline

\end{tabular}
\end{minipage}
\end{table*}

 For comparison two well known HI surveys have far looser criteria for identification: HOPCAT the identification catalogue associated with HIPASS done at Parkes, has made identifications up to 7.5 arc min off-source at 2000 $km s^{-1}$ (i.e. 65 kpc.), while ALFALFA, done with Arecibo, identifies galaxies up to 2 arc min off source at 8000 $km s^{-1}$ ( i.e. 70 kpc.). No wonder they are able to find an optical ID for virtually every radio source they find!  
 
 The above calculation [col 3 Table 1] actually yields a minimum for the expected number of spurious IDs because it ignores galaxies that could also be projected close to the source position from cylinders of radius $\hat{r}$ in front of the source and behind, each of depth h.
 The full calculation [col 4] is carried out in Appendix A entitled "Identifying galaxies with astronomical sources". It results in what we call 'The Ambiguity Equation' for the number M of spurious clustered galaxies lying within $\delta\theta'$ arc minutes of a source at a distance of d(Mpc.)$ \--$  \emph{ where it assumed  that the sources are clustered with the galaxies .}
 The Ambiguity :
\be M=\left(\frac{q\bar{n}}{22}\right)\cdot d(Mpc.)^{\frac{6}{5}}\cdot\left(\delta\theta'\right)^{\frac{6}{5}}\ee
 where q is the factor signifying how many other galaxies beyond $L_*s$ with conventional SBs would be admitted as identifications. For instance if the observer would be willing to accept galaxies up to 3 mag. fainter than  $L_*$ as plausible IDs then, for the SDSS Luminosity Function q would be $\simeq 4$ and if 4 mag q would be $\simeq 6$.

 If we apply (4) to the HIPASS/HOPCAT blind survey where $\delta\theta'$ =7.5 arc minutes then if q=6 the Ambiguity M:

  at a radial velocity of 2000 km.s$^{-1} M\sim$ 1.5

  at a radial velocity of 6000 km.s$^{-1}   M\sim $6

  and if we apply it to the ALFALFA survey at Arecibo :$\delta\theta'$ =2 arc min with q=6 the Ambiguity:

  at a radial velocity of 6000 km.s$^{-1}  M\geq $1

  at a radial velocity  of 18,000 km.s.$^{-1}   M\sim $4.5

 suggesting that both surveys are probably plagued by numerous misidentifications.( Why the larger dish is as imprecise as the smaller is explained in Section 4).

Exactly the same arguments apply to the identification of  QSOALS with galaxies. In that case it is more convenient to rewrite (4) as :
  \be d(Mpc.)\cdot \delta\theta'\leq \left(\frac{22M}{q\bar{n}}\right)^{\frac{5}{6}} \ee

  Damped Ly- $\alpha $ systems, which seem the most likely counterparts to galaxies, appear in the UV where IDs have to be made at redshifts beyond 0.25 (typically at 0.4 where d(Mpc.) $\sim $1300).Then, according to (5) reliable identification (q=6)  requires $\delta\theta'\leq0.03$ or about 2 arc sec, and yet observers confidently make IDs with $L_*$ galaxies \emph{several arc minutes away}. According to (5) , the Ambiguity at 2 arc minutes would be \emph{no less than 30}. Almost none can thus be considered reliable.

  In the simplest possible terms we require for identification a precision in the physical position [from(5)] of
  \be  \delta\bar{x}(kpc.)\leq\left(\frac{M}{q\bar{n}}\right)^{\frac{5}{6}}  \ee 
  which, for M=0.2 and q=6 $\simeq$ 11kpc. This is the severe challenge clustering poses the identification process.

  The simplicity of the above argument should at the very least cause  extra-galactic surveyors to pause for thought. Does it not place the burden-of-proof squarely on those who want to identify blind HI sources and QSOALS with prominent optical galaxies nearby? Could not the present consensus that dim and dark galaxies are rare be an entirely circular argument based on neglecting the tendency of galaxies to cluster so strongly?

 \section{RECENT VLA IDENTIFICATIONS}
 The uprated VLA or Karl Jansky Telescope (NRAO, 2015) finally has the  combination of sensitivity, bandwidth and resolution needed to detect dim and dark galaxies at 21-cm. and we have carried out a pilot program to look for them. But because integrations of $\simeq 2$ hours per field are needed there has to be some care in selecting the fields if the search is to be economical. We elected to examine  the fields of HIPASS sources whose published identifications appear, on probabilistic ground, to be suspect. We calculated Q for every source in the HOPCAT identification catalogue where Q is  the Odds on a particular identification being correct .
 Every source discrepant in position from a nearby optical galaxy is certainly not a dark galaxy. Radio positions have finite accuracy and what one needs to know is:
 
 \be Q(\Delta\theta)\equiv\frac{P(real)}{N(spurious)}\ee
 where P(real) $\equiv$ Probability that ID is correct but angular positional error is as great or greater than $\Delta\theta$ measured. And N(spurious) $\equiv$ number of plausible but spurious optical identifications which could be found within $\Delta\theta$ (see Sect. 2). When $\Delta\theta$ is small P(real) will generally be high, and N(spurious) $\--$ due to clustering, depending on the distance of the source from the observer, low, yielding high odds on a correct identification. But for larger $\Delta\theta $s, and larger distances out in space, Q will be small, suggesting that the true identification is possibly not the one chosen. A survey with many Q values below 1 could be plagued with  misidentifications, among which might eventually be found Dark Galaxies and Intergalactic Clouds.
 
 Simplistically P(real) will depend on the beam-shape of the radio-telescope and $\sigma$  the  Sig/Noise of the source. The positional accuracy possible on an unresolved source is usually reckoned to be
 \\

  $\delta\theta\sim \frac{HPBW}{S/N} \sim  \frac{\lambda/D}{\sigma}$
  \\

  As an example let us calculate approximate Q-values for the 4315 sources in HIPASS survey which were later identified with optical galaxies in the HOPCAT catalogue.  HIPASS which was carried out on the Parkes dish between 1997 and 2000, used 13 beams and two polarizations to find 4,315 sources in the Southern hemisphere. The identifications  were made in 'HOPCAT' (Doyle et al. 2005). HOPCAT used a combination of positional and radial-velocity information to pick its identifications. Positional matches up to 7.5 arc minutes, and velocity matches up to $\pm$ 400 km$s^{-1}$ were considered as 'reliable'. The strong impression was left that dark galaxies and intergalactic gas clouds must be rare or non-existent. To quote from the abstract: "Isolated 'dark galaxy' candidates are investigated using an extinction cut at $A_B$$ \leq$ 1 mag and the blank-fields category. Of the 3692 galaxies with an $A_{B}$ $\leq$ 1 mag, only 13 are also blank fields. Of these 12 are eliminated either with follow-up Parkes observations, or are in crowded fields. The one remaining one has a low surface brightness. Hence no isolated dark galaxies have been found within the limits of the HIPASS survey."

Q, on the contrary, reveals a very different picture[the details of how Q was calculated are in Appendix B]. About a quarter of the HOPCAT IDs have Q-values of less than 1, particularly those with angular discrepancies from their sources of 2 arc. mins. or more  and those out at radial velocities in excess of 2000 km.$sec^{-1}$. The strong claims made in HOPCAT that dim and dark galaxies are very rare or absent in the survey were based on the misunderstanding that the positions and velocities of galaxies are almost wholly independent $\--$ when the very  reverse is true. As we saw in Section 2 the Ambiguity of HOPCAT is between 1 and 6, implying that an optically bright but spurious counterpart could probably be found for most dim and dark galaxies. It should therefore be very well worth while to take a look at HOPCAT IDs with the VLA.

In the allocated time we managed to look at 17 HIPASS fields with questionable optical IDs as suggested by their low Q values(i.e. those in the lowest decile with Q's of 0.05 or less). We used the C configuration which in Natural mode has a beam-diameter of 25 arc sec. yielding the necessary physical resolution of 14 kpc. out at a redshift of 6600 $km. sec^{-1}$, the outer distance limit of our chosen sources. At that resolution we can expect that many of our candidate sources would be resolved so we had to select a combination of exposure time (roughly 2 hours), velocity-resolution ( 20 $km. sec{-1}$ when smoothed from an initial $ 2 km.sec^{-1}$) and HIPASS flux-density ($\geq 40 mJy.$) that would detect column densities per beam of $\sim 3\times\left(10\right)^{19}$ HI atoms cm$^{-2}$, typical of the outer discs of known HI galaxies.

We are still trying to obtain matching deep CCD frames but we can at least summarise the HI findings in so far as they bear on the Identification problem which is the main concern of this paper.

The 17 candidates,  have optical IDs as given by HOPCAT, at an average distance 5.3 arc min. away from the HIPASS radio positions, some at the permitted maximum distance of 7.5 arc min.. They have Ambiguities (q=6) of between 0.5 and 5.5 with a mean of 2.8.

In all fields the VLA found resolved 21-cm. sources close to the HIPASS radial velocity. Since the VLA fluxes of these sources all have S/Ns comfortable exceeding 10 the VLA HI positions should be accurate to better than 5 arc sec. thus misidentifications are more or less ruled out.

Of the HOPCAT optical IDs 8/17 were correct, 5/17 were half-correct (i.e. were one of 2 or more galaxies apparently contributing to the HIPASS signal), while 4 were definitely incorrect. Of these 4 incorrects two appear to be coincident with dwarf /LSB galaxies of a kind usually ignored by HOPCAT; they must have $ M_{HI}/L_{B} $ ratios in the tens in solar units. More interestingly 2 have no obvious optical counterparts in the DSS and yet have HI masses of several times $10^{9}$ solar masses, i.e.as much as the Milky Way, and are promising Dark, or at the least very Dim galaxy candidates (see Fig 1.). We are trying to get further observations to be more certain of their precise nature but already we can be certain  that the optical IDs published for the large HIPASS catalogue are not reliable enough to rule out a substantial population of Dim and Dark galaxies within it.

With its much increased band-width (i.e. velocity-range in this context ) the VLA ought to be a useful blind-HI survey instrument in its own right. Accordingly we searched the 17 data cubes between 500 and 8000$km.sec^{-1}$ for other sources beside those found earlier by HIPASS. We found 19 such Blind sources , all with S/Ns  comfortably above 10. Found between 1180 and 6270 $km. sec^{-1}$ they are necessarily weaker than the HIPASS sources, but have HI masses between $4\times \left(10\right)^{8} $ and $1\times\left(10\right)^{7}$ solar masses. 8 are accurately  centred on optical galaxies in the DSS, usually faint or dim ones. 9 have the same redshifts as HIPASS sources  to within  $\pm 200 km.sec^{-1}$ , suggesting they are members of the same Group. 12  have no optical counterparts in the DSS ( Nor in SDSS version DSR 12).

From the observations a very crude estimate of the volume-density of the blind sources can be arrived at because the volume of each of the 17 VLA beams out to 6000 km.sec$^{-1}$  is approximately 10 Mpc.$^{3}$. If we discard the 8 new blind sources with velocities close to  the already known HIPASS sources we find a volume density $\sim (19-8)/ (17\times10)\sim 6\cdot 10^{-2}$ sources $ Mpc^{-3} \sim$ the number of galaxies in the SDSS Luminosity Function down to 4 mag below $L_{*}$ [see Sect.2]. So the 'new' sources, if they are new, neither dominate, nor are they dominated by the already known population of galaxies.

In all then the VLA has has found 36 sources in 17 fields, (17 HIPASS and 19 Blind) of which 2+12 have no apparent counterparts at the level of the DSS while others appear to be associated with VLSBGs. For the present however we make no claims as to what these sources are because the existing DSS and SDSS optical data is simply not deep enough to be sure they are no more than ordinary dwarf galaxies, with in some cases high $ M_{HI}/L_{B} $ ratios. However the observations do raise serious doubts about the ability of single-dish blind HI surveys like HIPASS to set strong constraints on the population of Dim and Dark galaxies.  In Section 4 we will show that increasing dish-size unfortunately does not help. 

\begin{figure*}
\begin{center}
\includegraphics[width=6in]{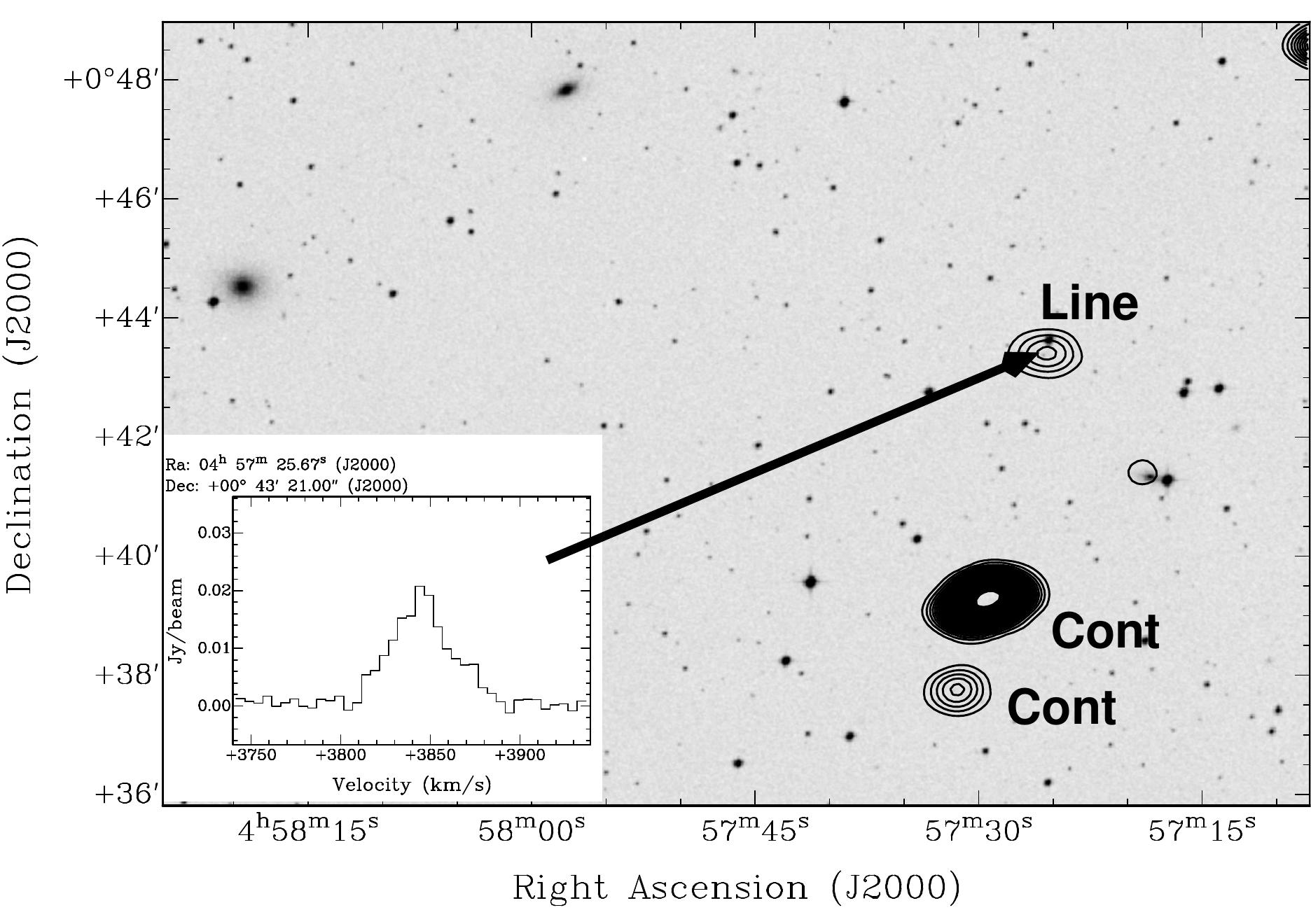}
\caption{ An example of a badly misidentified HIPASS source, HIPASS 0457+00. Our VLA HI contour map superposed on a DSS image, indicates the true position of the source (which is within 2 arc min. of the HIPASS radio position)  with an inlay of its HI spectrum which contains  2$\times10^{9}$ solar masses of HI. Presumably because of that large gas-mass  HOPCAT identified it with the bright early type galaxy 8.5 arc min. ENE at the top of the frame which in fact has no detectable HI at all in the VLA observations. There is no visible optical counterpart in the DSS image (The object visible within the HI contours is starlike and offset) but in the SDSS (version DR12) there is a faint (g=18.5) blue(g-r=0.15) extended( Petrosian radius 4.1$\pm$ 0.6 arc sec.) object right on the VLA centroid $\--$ which $\emph{could}$ be the core of a LSBG. It has as much gas as the Milky Way but is optically 100 times less luminous $\--$ exactly the kind of object we set out to find. On the basis of their Q-values (see above)  there could be hundreds of other HOPCAT misidentifications like this, hundreds of potential DDGs which have deliberately been found bright clustered optical counterparts which conform approximately to the HI radial velocity.In other words a strong bias against DDGs operated.}
\label{fig457.pdf}
\end{center}
\end{figure*}
\

                 \section{SINGLE DISH BLIND SURVEYS IN GENERAL}
                  
                  Since HIPASS there have been other multibeam surveys done with other single dishes, notably with Jodrell Bank and Arecibo. A larger telescope with a correspondingly smaller beam ought, one might suppose, be less affected by misidentification. But, as we shall now see, this is not the case because a larger telescope, having a smaller field-of-view, has to find its sources at a correspondingly larger distance away where its $\emph{physical}$ resolution, which is what counts for identification purposes, is no better. Indeed we shall reach a dismaying and surprising result: no single-dish telescope can carry out a clean  survey for dark galaxies and intergalactic gas clouds $\--$ clean in the sense that most of such detected sources will not possess plausible but spurious optical identifications. Because of strong clustering single dishes of any size do not possess the resolution requisite to their sensitivity. In other words their ability to detect HI sources at a distance has outstripped their ability to pinpoint the correct identification at that distance.
                  
                  In the usual notation the famous  'Radiometer Equation'  can be written (eg. Perley et al 1994)
                  
                  For detection: \be \frac{M_{HI}}{d^{2}\cdot \Delta V} > \frac{k\sigma T_{S}}{\sqrt{\Delta \nu \cdot t}} \cdot \frac{1}{A} \ee
                  where d is distance, A is collecting area, t is dwell-time and $T_{S}$ system temperature \-- the usual measure of noise, largely receiver noise in HI practice and $\Delta\nu$ is the bandwidth.
                  
                  Pointing at an $\emph{already known}$ point source the speed:
                  
                  \be \equiv \frac{1}{t } \sim  \left(\frac{A}{T_{S}} \right)^{2} \ee
                  
                  highlighting the advantage of antenna-size.
                  
                  However in a $\emph{blind}$ survey the dish has first to actually find the sources; then the smaller field-of-view of a bigger dish (diameter D) becomes a severe handicap. For a given source and S/N $\sigma$ the maximum distance you will find it will be, from (8):
                  \be d(max) \sim D \cdot t^{1/4} \cdot(M_{HI}/\Delta V)^{0.5} \cdot T_{S}^{-1/2}     \ee
                  
                  where $\Delta V$ is the line-width. The $\emph{survey\,speed}$ will be proportional to the number $\zeta$ of sources detected in survey time T where
                  
                  \be\zeta\propto [d(max)]^{3}  \cdot \Delta \Omega \cdot (T/ t ) \ee
                  
                  where $\Delta\Omega$ is the beam-area $\sim(\lambda/D)^{2}$. Thus survey-speed:
                  
                  \be\equiv\zeta/T \,\,\propto  N_{b} \, D\cdot t ^{-1/4}\cdot T_{S}^{-3/2} \ee
                  
                  where $N_{b}$ is the number of independent beams in a multi-beam receiver ( D=64m. 13 beams in HIPASS, t=450sec.). Thus a huge dish like Arecibo (D=305m.,7 beams , 40 sec) is only modestly faster (2.5 times)  for blind surveys than a smaller dish like Parkes.
                  
                  The speed of a survey increases with a shorter dwell-time t (see 10) because the dish skims across a larger area of sky picking out a closer and hence brighter selection of sources. But clearly there is a limit to such speed because, as we now demonstrate, the surface-brightness-sensitivity will eventually fall too low to pick up galaxies at all, and certainly the ones of lower column density:
                  
                  To see this assume the gas-cloud is larger than the beam (below) so that in the Radiometer Equation  
  $M_{HI} \rightarrow \Delta\Omega\cdot N_{HI}\cdot d^{2} $ while $\Delta \Omega \sim1/D^{2} \sim 1/A $  in which case, for detection:
  \be (N_{HI}/ \Delta V)\geq \frac{T_{S}}{\sqrt{t}} \ee
  
 i.e. the dish-size has dropped out. At first sight  puzzling this arises  because  the bigger dish sees a correspondingly smaller area of the sky (i.e. of HI) in its beam, which exactly cancels its advantage in collecting area. (Another way to look at it is to note that its receiver noise, which predominates, is projected onto a smaller area of sky and thus competes with a smaller HI signal,) And this surface-brightness effect shouldn't be underestimated in the present context. Huge diffuse galaxies such as Malin 1, which was larger than the NRAO 300 foot beam even out at a radial velocity of 25,000 km $s^{-1}$ , may be detectable only by longer dwell-time surveys. Likewise dwarf galaxies which, the evidence suggests, generally have lower surface-densities,  (Appadoo et al 2009),  may evade detection when they are resolved. Calculations and observations show that   dwell times $\sim 10^{3}$ secs. are needed to find objects with the mean column density $N_{HI}\sim 10^{20} cm^{-2} $and line-width$\sim 50 km.sec^{-1}$ of an object like the Milky Way which fills the beam. To find lower Surface brightness objects even larger dwell-times will be necessary.  Note that HIPASS had a dwell-time of 450 sec. Jodrell Bank 360 sec, [ but 3,600 sec. in the Virgo region], whereas ALFALFA has only 40 secs.
 
 Thus  in practice all blind surveys$\emph{ should}$ have a  minimum dwell-time/beam required by surface-brightness [i.e.  $N_{HI}/\Delta v$]  sensitivity. And because  of (10) $d(max) \sim D\cdot t^{1/4} $ a larger dish must find the majority of its blind sources correspondingly further away where its  $\emph{'physical'}$ resolution $\delta x$ will be no better:

   i.e. $\delta x \sim d(Mpc.)\cdot\delta\theta\sim Dt_{min} ^{1/4}\cdot\lambda/D\sim D^{0}$.

   Since some are puzzled by this result , assuming that larger dishes must be better for this purpose, we look at the matter in an alternative way via the Ambiguiity(4):

      $ M=\left(\frac{q\bar{n}}{22}\right)\cdot d(Mpc)^{6/5}\cdot\left(\delta\theta'\right)^{6/5}$
      which is a source-by-source estimate. The average Ambiguity:

      $\langle M\rangle =\left(\frac{q\bar{n}}{22}\right)\cdot\langle \left(d(Mpc.\right)^{6/5}\cdot \left(\delta\theta'\right)^{6/5}\rangle$

      which, since

 $d\sim Dt_{min} ^{1/4}$ and  $\delta\theta \sim \frac{\lambda}{D}$

      is independent of dish-size once again.

      In Section 2 we argued that  the HIPASS/HOPCAT blind survey was highly ambiguous, an assertion seemingly born out by our VLA observations of Section 3. We likewise calculated that the ALFALFA survey carried out with the much larger Arecibo dish is equally Ambiguous $\--$ and now we can see why. Because of their narrow beams large dishes used for blind surveys must find their sources proportionately larger distances away $\--$ where their $\emph{physical}$ resolution, and hence their ambiguity, will be no better.
      
      Larger dishes are unquestionably more sensitive and of higher resolution for studying sources $\emph{already found}$. But when it comes to finding blind sources for themselves most of their advantages vanish. No single-dish telescope can carry out a clean HI survey for dark galaxies and intergalactic gas clouds because, in the face of strong clustering, they do not possess the resolution requisite to their modern sensitivity.

In summary then, and irrespective of dish-size, the single dish surveys which set out to find dark galaxies, low surface-brightness galaxies and intergalactic gas clouds will probably have missed nearly all of them through a combination of poor physical resolution ( and therefore a strong tendency to misidentification because of clustering), and in some cases (e.g. ALFALFA) poor column-density-sensitivity because of short dwell-time.

\section{QSOALS AND HIDDEN GALAXIES}

Dim and even totally dark galaxies should nevertheless show up as absorbers in the foreground of distant ultraviolet sources such as QSOs $\--$ ultraviolet because most of the strongest absorbing lines lie in that region of the spectrum. The absorbers that might be attributable to whole galaxies ought to exhibit column densities higher than N(HI) $\ge 10^{20.3}$ HI atoms cm$^{-2}$, typical of local HI galaxies in emission. At such column densities the Lyman alpha lines are highly broadened by damping and therefore easy to find even in low resolution spectra, and some 5 to 10 per cent of high redshift QSOs show such DLAs (Damped Lyman-alpha Absorption systems) in their spectra(e.g. Prochaska et al 2005). There are many metal lines too so it is natural to suppose they originate in gaseous structures where star-formation is or has been going on i.e. galaxies. If those galaxies are at low enough redshift they might be visible and several large surveys have set out to find them using HST UV spectra as a starting point. The results were dramatic. Every such DLA could be associated with a giant galaxy no more than a minute or two of arc away, and with an almost identical velocity. Even so the impact parameters were so high $\--$ often 2 or 3 hundred kpc, that such giants were inferred to have colossal halos. The implausibility of such halos was discounted by the discovery of an anti-correlation between impact parameters and absorption-line equivalent-widths which seemed to argue for their existence, and seemed difficult to explain otherwise. With such gigantic halos all the absorbers could easily be accounted for by optically luminous galaxies, leaving no room for additional dim or dark galaxies. Thus in a review of the subject Lanzetta et al (1999) were able to confidently conclude that: " (1) most galaxies possess extended gaseous envelopes of $\simeq 160 h^{-1}$ kpc. radius, and (2) many or most Lyman-alpha absorption systems arise in extended gaseous envelopes of galaxies and that any 'unseen' low surface brightness galaxies are unlikely to contribute significantly to the luminosity density of the universe."

Alas such conclusions were based on spectral coincidences between QSOALS and galaxies in the redshift range 0.25 to 0.6 where, in view of clustering, the physical resolution required to make unambiguous Identifications ($\sim 11 kpc.$) translates into angular separations of less than 2 arc sec.  But QSOALS observers actually make  IDs at 1 to 2 $\emph{ arc minutes}$! They fell into exactly the same trap as we HI observers, failing to recognise that in a heavily clustered universe velocity and position are very far from independent. And note that the inferred giant halos have radii ($\sim 160 h^{-1} kpc.$) suspiciously close to the expected (projected) $\langle r\rangle $from any $L_{*} $to the nearest cluster galaxy capable of causing galaxy-like absorption (assuming there are q such per $L_{*} $ in the cluster) for then:
\\

$ \bar{n}q \int_{0}^{\langle r\rangle} 4\pi r^{2}\left[1+ \left(\frac{r_{0}}{r}\right)^{1.8}\right] \cdot dr \geq 0.5$
\\

yielding $\langle r\rangle$ = 250 kpc. for q=1, 180 kpc. for q=2 and 100 kpc. for q=3.

As for the clear anti-correlation observed between impact-parameter and absorption equivalent-width, e.g. Lanzetta et al. (1999), there is, under the clustering hypothesis,  a more natural alternative to giant halos as an explanation for that. A QSO sight-line passing through the inner volume of a Group will be, statistically speaking, closer to both the real dim-or-dark absorber and to the nearest big visible spiral chosen spuriously to be the absorber, than a sight-line passing through the outer volume of the Group. The observed anti-correlation is thus naturally expected without being physical (Linder S, 1999) and hardly  evidence in favour of giant halos. Indeed the high noise observed in that anti-correlation is more suggestive of the clustering explanation.

Recently QSOALS observers have become rather more cautious (e.g. Tripp 2008) though they are still devoting large HST and Keck resources to look for coincidences between QSO sight-lines and easily visible foreground galaxies  But Meiring et al 2011  reported that in their best studied case: "Follow up imaging with WFC-3 reveals several star-forming galaxies within 10 arc sec of (the QSO)......at first sight any of these candidates appears to be a promising candidate for the DLA-host. Follow up spectroscopy reveals that none of them are at the right z.......Even with this supporting data the origin of the DLA host remains enigmatic and exemplifies some of the difficulties of discovering the host galaxies of such systems." And in a more recent paper they admit (Battisti et al 2012) "The number of DLAs with spectroscopically confirmed hosts is very small." We are not surprised! The Ambiguity Equation (4)  alone suffices to show how extremely difficult this problem will be because it requires the elimination of possibly dim and dwarf absorbing galaxies $\emph{ a second of arc away from very bright}$ QSOs. Once a promising line of attack, this now looks more like a forlorn hope. Certainly QSOALS have so far set only weak constraints on dim and dark galaxies.

If $L_{*}$ galaxies don't have colossal halos then where do the great majority of QSOALS originate? If dim and dark galaxies are largely responsible instead, there must be a fairly rich population.

                          \section{DISCUSSION}
                          
                          The Conjecture "Most of the galaxies, even in our neighbourhood remain to be discovered because galaxies are extremely hard to detect through our atmosphere and against our sky."   is a natural one, with obvious implications for many areas of both astronomy and cosmology. It is also, epistemically speaking, a very healthy one because it both encourages exploration and invites refutation (Gauch, 2005).
                          
                          Broadly speaking there have been in recent years three approaches to examining the Conjecture, none encouraging and some positively damning. Optical observers have used increasingly more sensitive CCD detectors on increasingly larger telescopes to search in particular for Low Surface Brightness Galaxies $\--$with meagre results. Secondly Radio astronomers have made blind HI surveys, only to claim that virtually every source so detected can be associated with a bright optical galaxy nearby with, to within the combined errors, the same radial velocity. And thirdly QSOAL observers argue that  their absorptions can be explained if they assume that $L_{*}$ galaxies have colossal gaseous haloes 300 kpc. or more in diameter. If that is the case then there is no need or indeed room, for any further dim or dark absorbers. We have already cast doubt on these last two arguments, and will return to them later, but first we must tackle the failure of the deep CCD surveys (e.g. Driver et al .2005).
                          
                          It has been widely supposed that because very few LSBGs have turned up in deep CCD surveys carried out with 4-M class telescopes that therefore they must be rare but, we argue next, that  is not the case. Finding a dim object is a signal-to-noise problem, a matter of counting object-photons as compared to sky-photons. To find a really dim object it must therefore be of large angular size, i.e. be relatively near by to us in space. But nearby objects are relatively rare (e.g only 2 Magellanic Clouds) and so to find them a large area of sky must be surveyed to a considerable depth$\--$ an impracticability with a detector of small physical size like a CCD $\--$ despite its high sensitivity. The argument is not completely obvious and requires some algebra [ which we do in Appendix C] but once grasped is manifestly true. It reveals that  the number of galaxies of Luminosity L and surface-brightness I that will turn up in a survey lasting a total time T is:                           \\

           \be \hat{N} \propto \phi \cdot T\cdot \left(L \frac {I}{S}\right)^{3/2} \times \left[D^{3} t^{0.5} \right]\times  \{WQ^{3/2}\} \ee

           where  S is the surface-brightness of the sky, t the dwell-time per field, W the aerial on-sky FOV of the detector and Q its quantum efficiency, D the telescope diameter and $\phi $  the Luminosity-function.  The first bracket contains the properties of the galaxy, the second of the survey, the third of the detector.
           
           From ( 14) we infer the following:
           (i) To acquire a certain number of galaxies of SB I:
           \be t\propto \left( \frac{S}{I}\right)^{3} \ee
           thus, for a drop in surface brightness of 1 mag. the dwell-time t per field must be increased by 3 mag. or  a factor of 16. But the 'optimum 'Visibility' of galaxies, as we have long argued elsewhere (e.g Disney and Phillipps 1983) lies  within a narrow window of SB with a Full-Width-Full Maximum of only 2.5 mag. on the low-SB side [ see Disney and Lang 2012)]. Thus to escape entirely out of the window  to see  new populations of LSB galaxies we would need to increase the dwell-time t by 2.5 times 3 mag or a factor of one thousand!. We truly are imprisoned in a lighted cell. This far-from-obvious argument deserves to be more widely appreciated.
           
           (ii) The detector figure-of-merit $ WQ^{3/2}  $ is higher for CCDs than for Schmidt photographic plates [[36 sq. degs, Q $\sim 0.01 ]]$ provided the CCDs [[ Q$\sim 0.5$]] have $\geq$ 2000 pixels a side. The grasp of any survey is
\\

           $\propto T \times\left(D^{3} t^{0.5} \right) \times \left[WQ^{3/2}\right] $
           \\
           
           which means that  1-month-long CCD surveys with 4 metre-class telescopes will be an order of magnitude less effective for finding low surface brightness galaxies than the combined Schmidt surveys covering the whole sky done 30 years ago. However if photon counting were the whole story then the SDSS ought to beat the Schmidt surveys by a factor of 5 to 10, despite its very short dwell-time $\sim 100 secs$. Unfortunately, very low SB galaxies can only be detected if they look apparently very large [Appx C] when the unevenness of the sky background itself, not its photon statistics, becomes the predominant source of noise  ( Sabatini, Roberts and Davies 2003, Martinez-Delgado et al 2010) . Telescope size (D) may help but may run into serious pixel-matching problems as well as the aforementioned background fluctuations. Large arrays of CCDs, perhaps on medium-sized telescopes, might succeed eventually, but they will probably require observing strategies  and software dedicated to the task. For the moment the photographic Schmidt surveys of the 1950's to 1990's have, in this context, not been superseded. This is a sad admission because with a quantum efficiency of only one per cent, and saturation after an hour, they amount to no more than a 36 second glance at the universe.
           In summary then the hope we all had of finding a significant population of dim galaxies using deep CCD surveys has been frustrated by the need to find them nearby where large apparent size would be able to overcome low signal-to-noise per unit area. This frustration was not obvious and would have been difficult to recognise before the advent of linear panoramic detectors.

                          The failure of the 21-cm attempts to refute  the Conjecture we have discussed. The earliest such attempt, using Arecibo locked in transit mode (Zwaan et al ,1997) ,was ingenious but  seriously ill-found (Minchin R F et al 2003).  An internal comparison of their own data suffices to reveal this. If their data were as deep as they supposed, how could they pick up virtually all of their 66 sources in brief snapshots ( 20 mins) with the smaller VLA?  And why was their source-count at least an order of magnitude too low? One suspects that they somehow failed to accumulate their daily scans, leaving a shallow single-transit survey setting few interesting limits.
                          
                          The failures with HOPCAT  we have admitted and the weaknesses of  ALFALFA  we have pointed out. Both are faults of misidentification based on the plausible but fallacious idea that, in a highly clustered universe, positional and velocity information are more or less independent. Our VLA observations confirm only too well the outcome of the clustering calculations. Up to a quarter of the IDs we looked at proved to be wrong; the incorrectly identified sources appear to be associated with faint, dim and even blank fields at the level of the DSS. Moreover of the 19 $\emph{new} $ blind sources which turned up in those same fields 12  appear to have no obvious optical counterparts.
                          
                           The misidentification 'hypothesis' also clears up the  otherwise serious problem for blind HI surveys; what we call  'The Inchoate Galaxy Problem' (Disney 2008, Appadoo et al 2009). Inchoate Galaxies are extremely low surface brightness (barely apparent in SDSS) dwarf objects, generally blue[(B-V) as low as 0.3] embedded in large amounts of neutral Hydrogen [$M_{HI}/L_{B} \geq  5$ in solar units.] $\--$ the prototype being the so called 'proto-galaxy' found serendipitously by Giovanelli and Haynes (1989). The Inchoate label for these objects derives from their apparent total lack of organisation. More irregular than Irregulars they have no cores or obvious centres, and appear as merely haphazard enhancements of surface brightness at what appear to be HII regions.
                          
                          The dozen or so 'Inchoates' within our Equatorial Survey (Appadoo et al 2009, Disney et al 2008) of some 200 galaxies presented an inexplicable $\emph{combination}$ of properties. Their high gas fractions indicate little integrated past star formation, while their blue colours can only be explained by a recent burst of star formation. Either they are young$\--$ in which case where are their totally dark HI predecessors? Or they are briefly bursting and will quickly fade away $\--$ in which case where are their faded or totally darkened remains? Neither kind of object appears in HIPASS, an identical survey but with a different identification program. Hence the mystery. But if many HIPASS sources are misidentified with brighter galaxies, all is explained(see Disney and Minchin 2003 for an alternative but now unnecessary  explanation in terms of 'frozen hydrogen').
                          
                           This last argument can be reversed.The very existence in blind HI surveys of Inchoate Galaxies which, because of their blue colours, individually must have comparatively short lives, requires there to be a larger population of optically invisible HI clouds $\--$ either the antecedents and/or the decendents of those that are presently observed to contain a predominant population of evanescent blue stars. That they are not presently detected suggests there is something seriously amiss with the existing identification process .
                           \\
                           \\

                            What  of all the work that has subsequently been done on sources picked up in blind HI surveys and for which the identifications must now be so doubtful? Caveat emptor. We will speak here only of the Equatorial Survey which we carried out ourselves with the Parkes Multibeam as part of the HIPASS survey, and which covered  2  steradians and found 1100 sources (Appadoo et al 2009). The part published comprises the 1700 sq. degs. which overlapped the SDSS DR-2 release. Originally that subset comprised 370 sources which we identified optically, using almost identical methods to HOPCAT. However we then threw out the 175 most doubtful IDs leaving only  195  sources in our statistical analysis . The $\emph{identified} $ source density is $\sim$ 195/0.6 $\simeq 300$ sources/steradian ( a factor of 3 down on HOPCAT). Comparing that fraction with our calculated  Q values  suggests we must have thrown out most of the bad apples.  There probably still are a handful, as we admitted at the time, but not enough to seriously challenge the largely statistical conclusions (Disney et al 2008).

.
                 The  other strong refutation of the Conjecture , claimed by the QSOALS community is, if anything , more incredible  $\--$ but for the same basic reason. Identifications were made with bright spirals typically 160 kpc. away from the QSO line-of-sight (Sect 5). But 160 kpc.  is almost exactly the average distance from an $L_{*}$ galaxy to the nearest dwarf or dim galaxy in the Group capable of causing the absorption.  This is highly suspicious, and not at all supportive of a physical association, especially not one implying colossal halos without much other evidential support. Those claims were mainly buttressed by the clear anti-correlation observed between impact-parameter and absorption equivalent-width e.g. Lanzetta et al. (1999). However there is an alternative explanation for that under the clustering hypothesis: a QSO sight-line passing through the inside of a Group will be, statistically speaking, closer to both the real dim-or-dark absorber and to the nearest big visible spiral chosen spuriously to be the absorber, than a sight-line passing through the outer part of the Group. The observed anti-correlation is thus naturally expected, without calling for any physical link(Linder S, 1999). The whole case for Giant Halos, which is venerable (Bahcall and Spitzer, 1969), needs to be re-examined.

                              It is notable that QSOALS observers first, then we blind-HI-observers second, fell  into the same egregious trap. Setting out to $\emph {look for}$ optical counterparts we, thanks to clustering, easily found visible galaxies nearby. Then, when we checked the radial velocities  $\--$ Lo, they matched. It was all very tempting and very natural to claim an identification . But had one started with  a different mind-set $\--$ to look out for dim and dark galaxies $\--$ then most such tempting matches would be regarded sceptically, or dismissed out of hand. It is, like so many debates, a burden-of-proof issue. In the context of ruling out dim and dark galaxies the burden of proof here should surely  lie on those who seek to claim identifications between sources and plausible visible objects in their neighbourhoods. If that is conceded, the existing confusion melts away. Likewise the failure of optical observers to find dim galaxies with CCDs on large telescopes. Had we realised the odds were all against us we would have understood that the lack of positive evidence was not significant evidence of a negative kind. Any would-be explorer of the unknown must nail the burden-of-proof issue first.

                            The HI data, the QSOALS data and the failure of optical observers to find LSBGs, all seemed to provide cohering evidence against the existence of Hidden Galaxies and Intergalactic Gas Clouds in any numbers. And yet, as we can now see, none of those claims stands up to examination. \emph {That is why it is necessary to be outspoken here}. What only matters is that The Conjecture is  still very much alive. There may well be a Low Surface Brightness Universe out there rich in both information and material. It may prove tricky to find, so the last thing needed is any false prejudice against its existence based on claims that are no longer credible. Absence of evidence is certainly not evidence of absence here.
                            
                            An example of such prejudice at large was the storm of criticism (see Cho A,  2007), especially from the 'expert' HI community, directed at the one really plausible Dark Galaxy candidate that turned up in the blind HI surveys, namely VIRGOHI21 (Davies JI et al 2004, Minchin et al 2005, Minchin et al 2007). Impressed perhaps by their own ability to find optical counterparts for every other HI source in the sky, they dismissed it as 'tidal debris' (e.g. Haynes et al 2007),  $\--$ which it could not be, simulated it away (Bekki et al 2005, Duc and Bournand 2007 ) or proved it could never have formed in the first place (Taylor and Webster 2005). It is time to reconsider the favourable evidence $\--$ which we contend is very   strong (Minchin et a. 2007) $\--$ . The point is that no dynamical simulation has been able to model the unique and abrupt change of gas velocity observed (~ 200 km.$s^{-1}$ within 16 kpc.) without invoking a massive ($> 10^{10} $ solar masses) dark object close to the line of sight (Vollmer, Huchtmeir  and Van Driel  2005;  Minchin et al 2007).
                            \\

                            We have so far only attacked the evidence against the Conjecture as embodied in certain probable misidentifications; we cannot leave without at least mentioning the strong circumstantial evidence in its favour, evidence discussed at far more length in two recent IAU conference proceedings: 'The Low Surface Brightness Universe '(ed Davies J I et al 1999) and "Dark Galaxies and Lost Baryons"(ed Davies JI and Disney MJ 2008).  But it is hard, very hard to ignore. There are so many independent coincidences in galaxy photometry $\--$ half a dozen at least (Ellis GFR et al. 1984; Disney,1999) $\--$ that are explained if the The Conjecture is true, that are a complete mystery if it is not: (a) Optically discovered disc galaxies with a wide range of luminosities have surface brightnesses(SBs) which cluster around a  peak value (Freeman 1970, Disney 1976. (b) The peak value is just such as to give discs the most prominence against the terrestrial sky (Disney 1976). (c) Optically discovered ellipticals  with a wide range of luminosities have SBs which cluster around another peak value (Disney, 1976). (d) That peak value, very different from (b), is  again just such as to render ellipticals most prominent against the earthly sky (Disney 1976) . (e) There is a complete lack of correlation between Blue SB and blue  (B-V) colour amongst discs $\--$ when increased star-formation should push both parameters up together (Disney and Phillipps 1985, Bothun, Impey and McGaugh 1997. (f) A lack of correlation is observed between the apparent SBs of galaxies and their galactic latitude $\--$ even in latitudes where the foreground absorption ought to be significant (Davies JI et al 1993). (g) There is a close correspondence between the calculated Visibility  (i.e the volume within which they can be detected) of galaxies and their median observed distances in a large sample of spirals (Davies JI et al 1994). And not the least is (h) the extraordinary discovery that virtually all the galaxies in the Hubble Deep Field have the same apparent Surface-brightness, irrespective of their redshifts and the same as galaxies nearby (Jones and Disney 1997, Disney and Lang 2012) . But HDF redshifts range from $\leq$ 1 to 8 and surface brightnesses ought to vary as the Tolman factor $(1+z)^{-4}$  or by 10 magnitudes! What we are probably seeing out there is the very high surface-brightness population of galaxies which we miss from the ground, because of strong selection effects. We could just as well be missing the low surface brightness population as well.  A large helping of extra ( i.e. hidden) galaxies at high redshift is exactly what we need to explain the re-ionisation of the universe (e.g. Robertson et al 2010).

\section{ACKNOWLEDGEMENTS}

MJD would like to thank Dr Helene Courtois from the University of Lyon and Dr Valentina Karachentseva from the Astronomical Observatory of the University of Kiev, who concentrated his attention on the serious misidentification problems there must be with the HOPCAT catalogue  during a conference on Dwarf Galaxies at the Special Astronomical Observatory of the Russian Academy of Sciences, Nizhny Arkhyz, Karakai-Circassia, in 2009.
\\

The National Radio Astronomy Observatory is a facility of the National Science Foundation operated under a cooperative agreement by Associated Universities Inc.

\begin{appendix}
\section{IDENTIFYING GALAXIES WITH ASTRONOMICAL SOURCES}
 Rather general methods for source identification  have been published by a number of authors (eg Sutherland and Saunders 1992, Ciliegi et al., 2003) however they result in general expressions with integrals over rather general functions of variables which include angular distance from the source, magnitude, and morphological type. Thus it is difficult to draw conclusions simply and transparently from them without carrying out numerical integrals. Here we derive instead a simple analytical expression the "Ambiguity Equation" for the number M of spurious cluster galaxies lying within $\delta\theta\prime$ arc mins. of a source at a distance of d(Mpc.), $\emph{where it is assumed that the sources are} $ \\$\emph{clustered with 
 the galaxies}.$
 
 The Ambiguity \be M=\left(\frac{q\bar{n}}{22}\right)\cdot d(Mpc.)^{\frac{6}{5}}\cdot\left(\delta\theta\prime\right)^{\frac{6}{5}}\ee

         where $\bar{n}$ is the volume density of $L_{*}$ galaxies $Mpc^{-3}$ in the search area, and q is the factor by which $\bar{n} $ should be multiplied to take account of other galaxies beside $L_{*}$s with conventional surface brightnesses that would be admitted as identifications. For instance if the observer was willing to accept galaxies up to 3 mag fainter than $L_{*}$ as plausible IDs then, for a typical optical luminosity function (e.g. Blanton et al) q would be $\sim 4$, and if 4 mag q would be $\sim 6$.                             
                                                        
                            To prove (A1) we need to calculate the number of galaxies expected in a cylinder of radius r' and depth 2h (a distance h in the foreground of the source, and h behind) where r' is the acceptable positional error in $\emph{physical}$ distance projected onto the sky[ = $\pi /4 \times$ true distance r on average]. Assume the clustering is described by the well known clustering function $\xi(x)$ such that the number expected in a sphere of physical radius r surrounding the source is
                            
                            \be \eta(r)=\bar{n}\cdot \int_{0}^{r}4 \pi x^{2}\cdot dx \cdot [1+\xi(x)] \ee
                            
                            where \be \xi(x) = (r_{0} /x)^{1.8}        \,\,\, (r_{0} = 8 Mpc) \ee
                            
                            e.g. see Peebles (1993) and $\bar{n}$ is the average number-density of relevant galaxies.
                            
                            To  the above number in the sphere must be added the numbers in the two end cylinders stretching from r' to (r'+h) beyond the source and from -r' to -(r'+h) in the foreground. These will add
                            
                            $\eta $(both cylinders, projected radius r')
                            
                            \be \approx 2\pi \bar{n} r'^{2}\int_{r'}^{h} dh \cdot [1+\xi(h)] \ee
                            
                            where the approximation arises from assuming that h$\gg$r'. [In the case of 21 cm sources h should be $\Delta V/H_{0}$ where $\Delta V$ is the maximum permitted velocity discrepancy between source and ID which, for HI sources at least, has generally been set at $\ge$ 100 km s$^{-1}$. Thus the cylinders are of order 1Mpc    deep as  compared to tens of kpc for r' $\--$ validating the approximation in the HI case.]
                            
                            Evaluating (A4):

    \be\eta(cylinders) \approx2.4\pi\bar{n} (r')^{1.2}\cdot(8)^{1.8}\cdot\left[1-\left(\frac{r'}{h}\right)^{1.8}\right]\ee
                            
                            and the last term will be negligible compared to 1 (implying that the far ends of the cylinders are barren of $\emph{clustered}$ galaxies.)
                            
                            With these approximations:
                            
                            $\eta(r')\approx 10\cdot \bar{n} \cdot (r')^{6/5}\cdot(8)^{9/5}$
                            
                            $\eta(cylinders,r')\approx \, 8\cdot  \bar{n}\cdot (r')^{6/5}\cdot (8)^{9/5}$
                            
                            Thus \be\eta(clustered)\equiv \acute{M}\approx 18\bar{n} (r')^{6/5}\cdot (8)^{9/5}\ee
                            
                            which can be compared with the number of unclustered  galaxies in the whole structure:
                            
                            \be\eta(unclus) = 2\pi\bar{n}(r')^{2}\cdot h   \,\,,\,\, [h=\frac{\Delta V}{H_{0}} ]\ee 
                            
                            Comparing (A6) with (A7):
                            
                            $\eta$(clustered)/$\eta$(unclustered)$\approx (18/2\pi)\cdot (8)^{9/5}/\left(\frac{\Delta V}{H_{0}}\right)\cdot (r')^{4/5}$
                            
                            which, for any practical combination of $\Delta V$ and r' (21 cm) is of order a thousand or more. It is therefore true, as claimed, that virtually every plausible identification with an HI source will be with a member of the same Group or cluster. To all intents and purposes, and unless the velocity information is very precise, (i.e with combined optical-minus-radio errors of better than 15 km $s^{-1}$) the velocity information is $\emph{not}$ independent  of the positional information; the two will be very closely correlated. [It was overlooking this tight correlation that led to over optimism in the identification process].

                            We can bring in the angular positional error $\bar{\delta\theta}$ (radius) because
                            
                            \be r'=d(max)\cdot \bar{\delta\theta}\ee
                            
                            where d(max) is the distance out to the source. Then if we convert $\delta\theta$ into arc mins. (A6) becomes:
                            
                            $M=\frac{1}{22}\cdot\bar{n}\cdot \left(d(Mpc.)\right)^{6/5}\cdot \left(\delta\theta'\right)^{6/5} $

                            Then introducing q (above):

                             $M=\left(\frac{q\bar{n}}{22}\right)\cdot d(Mpc.)^{\frac{6}{5}}\cdot\left(\delta\theta'\right)^{\frac{6}{5}}$

                             which is (A1).

                             M rises only slowly with $\delta\theta$ because of the strong clustering.

                             Examples: 

(a) If we apply (A:1) to the HIPASS/HOPCAT blind survey when $\delta\theta'$ =7.5', q=6 and $\bar{n}=10^{-2} gals.Mpc.^{-3}$ then ($H_{0}$ = 75 throughout):
                             
                             at radial velocity = 6000 $km.sec^{-1} \  \ M\sim  6$

                             at radial velocity=  2000 $km.sec^{-1} \  \ M\sim 1.5$

                                               (b) And if we apply(A:1)  to the ALFALFA blind survey at Arecibo  $\delta\theta'$ = 2'

                             at radial velocity = 18,000 $km.sec^{-1} $ \ \ M$\sim$ 4.5

                             at radial velocity = 6000 $   km.sec^{-1} $\ \ M$\geq 1$

                             So both surveys must be plagued by numerous misidentifications.
                             
                             A useful alternative is to write (A:1):
                             
                             \be d(Mpc.)\cdot\delta\theta' \leq\left(\frac {22M}{q\bar{n}}\right)^{5/6} \ee
                             If we use (A;9) to look at QSOALS then for mostly eligible identifications (i.e. M$\leq$ 0.2; q=6) and considering most such IDs have to be made at redshifts $\geq$ 0.25 (typically  0.4 where d(Mpc.)$\simeq$ 1300 Mpc ) requires $\delta\theta'< 0.03'\sim $ 1.5 arc sec. ,and yet observers confidently make identifications with galaxies $\emph{several arc minutes away}.$ According to the Ambiguity Equation the Ambiguity M at 2 arc. min. would be $\emph{no less than 30}$. Almost none of those QSOALS Identifications can be considered as reliable.
                             
                             Finally, in the simplest possible terms, we require a precision in the $\it {physical}$ position of:
                             
                             \be \delta\bar{x} (kpc.) \leq 4\left[\frac{M}{q\bar{n}}\right]^{5/6} \ee

                             which for M= 0.2, q=6 is 11 kpc.. This is the very severe challenge clustering poses to the galaxy identification process.

\section{CALCULATING Q-VALUES FOR THE HIPASS/HOPCAT SURVEY} 
Simplistically P(real) will depend on the beam-shape of the radio-telescope and $\sigma$  the  Sig/Noise of the source. The positional accuracy possible on an unresolved source is usually reckoned to be

  $\delta\theta\sim \frac{HPBW}{S/N} \sim  \frac{\lambda/D}{\sigma}$

  As an example let us calculate approximate Q-values for the 4315 sources in HIPASS survey which were later identified with optical galaxies in the HOPCAT catalogue.

 HIPASS which was carried out on the Parkes dish between 1997 and 2000, used 13 beams and two polarizations to find roughly 4,000 sources in the Southern hemisphere. It's positional uncertainties have been discussed at some length by Meyer et al (2003), Zwaan et al (2003) and Koribalski et al (2004 ). For sources smaller than the beam ($\sim$ 15 arc min) the positional errors are approximately Gaussian with a standard deviation $\sim$ (Half-power-beam-width)/(S/N). In practice this translates to

\be P(real)=4\left[1-erf \left(\frac{\theta}{\sigma_\theta}\right)\right]\ee

where    $\sigma_\theta=\frac{7.1}{S_{int}}$ +0.6  arc min and $S_{int}$ is the integrated flux for the source in Flux units-$km.s^{-1}$.
For closer sources larger than the beam (B1) is unsuitable. As a general rule it is found that HI sources have much the same column density, implying that source diameters grow with integrated fluxes and in the case of HIPASS Meyer et al (2003) found

\be S_{int}\simeq1.2\theta_{HI}^2\ee

where $\theta_{HI}$ is the diameter of a galaxy at the 1 solar-mass/$pc^{2}$ $(1.3\times 10^{20}$ HI atoms $cm^{-2}$) HI contour level in arc minutes. Thus bright sources will have large apparent sizes and the uncertainty in their measured positions will reflect this. If we assume that the probability of the measured position of the source being discrepant from its centroid is
\be P(>\theta)\simeq\frac{amount\,of\,HI \, beyond\,\theta}{all \,HI \,in \,source}\ee

and as HI is distributed approximately exponentially in most sources ,it is easy to show that

\be P(>x)\simeq e^{_x}(1+x)\ee
where $x\equiv \theta /\theta_{\beta}$

and $ \theta_{\beta}=0.2\sqrt{\frac{S_{int}}{1.2}}$

because, in any exponential disc, 90 per cent of the material lies inside x=4, i.e. 4 HI scale- lengths $\theta_{\beta}$.

The crossover point for the two kinds of positional error occurs close to
$S_{int} \simeq 10 Flux-units-km s^{-1}$.

i.e for $S_{int} \geq 10$ use (B3), otherwise use (B1).
\\
\\
In a multi-beam survey a source will tend to pass through or between several beams and its catalogued position will be some weighted mean of several beam signals. Thus a gaussian distribution for P(Real) [ (B1) above] can only be an approximation unlikely to be accurate out in the wings where P(Real) (gaussian) would be unrealistically small . We therefore regard Q as a useful tool for \emph{ranking}  the quality of IDs rather than an accurate estimate of P(Real) (This was born out by the VLA observations). About a quarter of all HOPCAT IDs have Q values of less than 1  while a high proportion  of those with angular discrepancies from their sources of 2 arc min. or more or those out at radial velocities in excess of 2000 km. $sec^{-1} $ have fractional Q values. For our VLA observations we chose only sources with Q values in the lowest decile ( i.e. with values of less than .05).

 \section{DETECTING LOW-SURFACE BRIGHTNESS GALAXIES:THE SIGNAL-TO NOISE PROBLEM.}
 If we have a galaxy image $\Theta$ pixels across, of SB I, where we have collected an average p photons/pixel, then the signal-to-noise $\sigma$ of the image will be
 \be \sigma\sim \Theta^{2} p\left(\frac{I}{S} \right)/\sqrt{\Theta^{2}p}\sim \frac{I}{S} \Theta\sqrt{p}\ee
 
 where S is the sky-brightness, assumed brighter than that of the galaxy, and the main noise is assumed to be photon-noise. Now
 \be \Theta\sim R/d\sim\sqrt{\frac{L}{I}}/d \ee
 i.e. $\Theta $ will be larger for a nearer galaxy   $\--$ this is obvious but crucial.(R is the physical radius, L the Luminosity)

.
 Now the number $\hat{N}$ detected in a survey of length T with a dwell-time per field t, using a detector of solid-angle $\Omega$ projected on the sky, will be ;
 \be \hat{N} \propto \frac{1}{3} \Omega \cdot d^{3}\frac{T}{t}  \phi \sim \frac{1}{3}\phi \Omega \frac{T}{t} \left(\frac{\sqrt{\frac{L}{I}}}{\Theta}\right)^{3}\ee
 
 where $ \phi $ is the `luminosity function and d is the outer boundary of the survey.

 But (C:1):
 
 $ (  )^{3} =\left(\sqrt{\frac{L}{I}}\cdot\frac{I}{S}\cdot 1/\sigma 
 \cdot\sqrt{p}\right )^{3}$

$ =L^{3/2}I^{3/2}S^{-3}\cdot \sigma^{-3}\cdot p^{3/2}$

 while
 
 $p\sim D^{2}Q S t  $  (D = telescope diameter, Q is Quantum efficiency,) so

 $(  )^{3}\propto L^{3/2}I^{3/2}S^{-3}\cdot \sigma ^{-3}\cdot D^{3} Q^{3/2
 } S^{3/2} t^{3/2}$

\be \sim L^{3/2} I^{3/2}S^{-3/2}D^{3} Q^{3/2} S^{3/2}t^{3/2}\sigma^{-3}\ee
 Substituting this last in (C:3) :
 \be\hat{N}\propto\phi \cdot T\cdot  \left(\frac{LI}{S}\right)^{3/2} \times \left[D^{3}t^{0.5}\right] \times \{\Omega Q^{3/2}\}\ee

  which is (14) in the main text.

\end{appendix}
  
\end{document}